\documentclass[twocolumn,showpacs,preprintnumbers,amsmath,amssymb]{revtex4}

\usepackage{graphicx}
\usepackage{dcolumn}
\usepackage{bm}

\begin{document}

\preprint{APS/123-QED}

\title{Uncovering collective
listening habits and music genres \\
in bipartite networks.}

\author{R. Lambiotte}
\email{Renaud.Lambiotte@ulg.ac.be}

\author{M. Ausloos}
\email{Marcel.Ausloos@ulg.ac.be}

\affiliation{%
SUPRATECS, Universit\'e de Li\`ege, B5 Sart-Tilman, B-4000 Li\`ege, Belgium
}%

\date{09/07/2005}

\begin{abstract}
In this paper, we analyze web-downloaded data  on
people
sharing their music library, that we use as their individual musical signatures (IMS).
The system is represented by a bipartite network, nodes being the music groups and the listeners.
Music groups audience size behaves like a power law, but the individual music library size is an exponential with deviations at small values. 
In order to extract structures from the network, we focus on correlation matrices, that we filter by removing the least correlated links. This
 percolation idea-based  method reveals the emergence of social communities and music genres, that are visualised by a branching representation. 
Evidence of collective listening habits that do not fit the neat usual genres defined by the music industry
indicates an alternative way of classifying listeners/music groups. The structure of the network is also studied by a more refined method, based upon a random walk exploration of its properties.
 Finally, a
personal identification - community imitation
model (PICI) for growing bipartite networks is outlined, following Potts ingredients. Simulation results do reproduce quite well the empirical data.
\end{abstract}


\maketitle

\section{Introduction}

Answering a common question such as "What kind of music do you listen to?" is not an easy task and is full of hidden informations about oneself.  Indeed, music is omnipresent in our society and is part of everyday life.
Moreover, it is well-known in social sciences  \cite{everyday} that music does not function merely as entertainment, but is deeply related to identity-building and community-building. In that sense, personal musical choices derive from a subtle interplay between cultural framework inheritance,  social recognition and personality identification. These reinforce one's self-image and send messages to others  \cite{renfrow}.
Due to the complexity of taste formation and the richness of available music, it is tempting 
to postulate that someone's music library is a unique signature of himself  \cite{bull}. For instance, it is interesting to point to the empirical study by D'Arcangelo \cite{gideon1} which shows that listeners strongly identify with their musical choice, some even going so far as to equate their music collection with their personality:
{\em My personality goes in my iPod}, as an interviewed person claims.
Consequently, it is difficult for people to recognise themselves in usual music divisions, such as punkers versus metal heads, or jazz versus pop listeners.
And, more commonly, they
answer to the above question "Everything... a little bit of everything".

Recently attempts have been made to characterise the musical behaviour of individuals and groups 
using methods from quantitative sociology and social network analysis.
These attempts were made possible because of the huge amount of music databases available now, associated with the current transition from materialised music (LP's, CD's...) to computer-based listening habits (iTunes, iPod...).
Amongst other studies, let us cite 
 the recent empirical work by Voida et al.  \cite{voida}, which shows that people form judgements about colleagues based on the taste - or lack of taste - revealed by their music collection, and admit to tailoring their own music library to 
project a particular person.
 
 The present paper focuses on these musical behaviours from a statistical physics 
 and statistical point of view, 
by analysing individual musical signatures and extracting collective trends.
This issue is part of the intense ongoing physicist research activity
on opinion formation \cite{holyst,galam,sznajd,castellano,kuperman}, itself
related to phase transitions and self organisation on networks \cite{xxz,xzz,bbb}, including clique formation \cite{neil}.
The characteristics of such phenomena depend on the type of network, as well as on the data size, thereby questioning universality, in contrast with Statistical Mechanics.

In section 2, we extract empirical data  from collaborative filtering websites, e.g. {\em audioscrobbler.com} and {\em musicmobs.com}. These sites possess huge databases, that characterise  the listening habits of their users, and allow these users to discover new music.  Our analysis consists in applying methods from
complex network theory  \cite{albert} in order to characterise the musical signatures of a large number of individuals. In section 3, we
present original percolation idea-based (PIB) methods in order to visualise the collective behaviours.  
They consist in projecting the bipartite network, where listeners and music groups are linked, onto a unipartite network, i.e. a network where listeners are connected depending on the correlations between their music tastes. Let us stress that  the usual projection method (\cite{newman0}, see details below), used in co-authorship networks for instance, does not apply in the networks hereby considered, as it leads to almost fully connected networks.
In this work, we also project the bipartite network on a network of music groups, and probe the reality of the usual music divisions, e.g. rock, alternative $\&$ punk, classical. We propose a quantitative way to define more refined musical subdivisions. These sub-divisions, that are not based upon usual standards but rather upon the intrinsic structure of the audience, may lead to the usual music genres in some particular case, but also reveal unexpected collective listening habits. Let us note that other techniques may also lead to an objective classification of music, e.g. by characterising their time correlation 
properties \cite{ivan}. In general, the identification of \textit{a priori} unknown collective behaviours  is  a difficult task \cite{vicsek}, and of primordial importance in the structural and functional
properties of various networked systems, e.g. proteins
\cite{ravasz}, industrial sectors \cite{onnela}, groups of people \cite{watts}...
Consequently, we also use another method in section 4 in order to uncover these structures, i.e. the percolated island structure is explored by randomly walking (RW) the network, and by studying the properties of the RW with standard statistical tools.
Finally, in section 5,  we present a growing network model, whose ingredients are very general, i.e. personal identification and community imitation (PICI). The model reproduces the observed degree (number of links per node) distributions of the networks as well as its internal correlations.

\section{Methodology}

Recently new kinds of websites have been dedicated to the sharing of musical habits. These sites allow first members to upload their music libraries, previously stocked on their computers, towards a central server, and next to create a web page containing this list of music groups. Additionally, the website proposes the users 
to discover new music by comparing their taste with that of other users. 
These methods of making automatic predictions for the interests of a user by collecting information from many (collaborating) users  is usually called {\em  Collaborative Filtering} \cite{wiki}. 
The data that we analyse here has been downloaded from {\em audioscrobbler.com} in January 2005. It consists of a listing of  users (each represented by a number), together with the list of music groups the users own in their library. This structure directly leads to a bipartite network for the whole system. Namely, it is  a network composed by two kinds of nodes, i.e. the persons, called users or listeners in the following,  and the music groups. The network can be represented by a graph with edges running between a group $i$ and a user $\mu$, if $\mu$ owns $i$.  

\begin{figure}

\hspace{-0.7cm}
\includegraphics[angle=-90,width=3.60in]{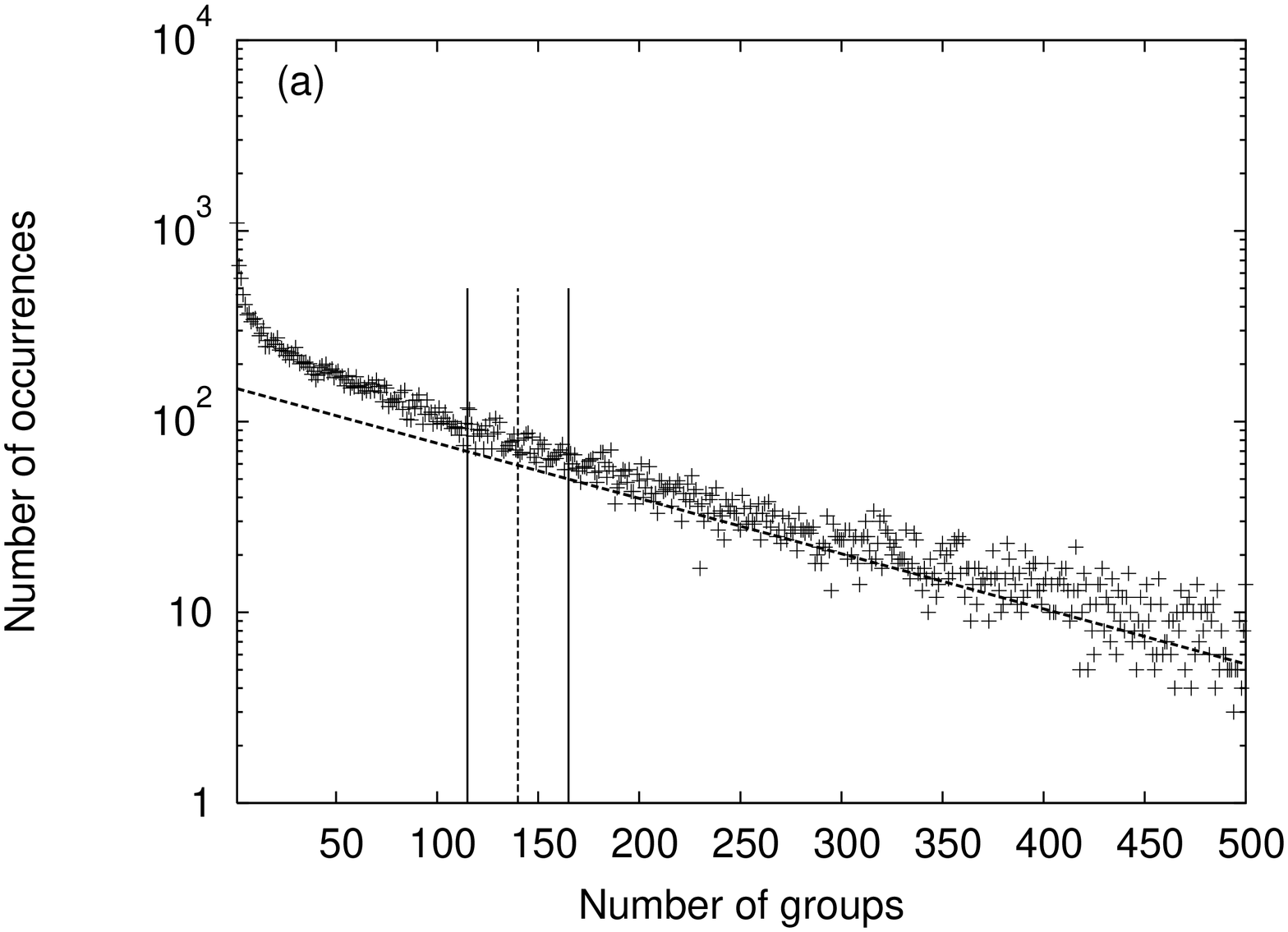}

\hspace{-0.7cm}
\includegraphics[angle=-90,width=3.60in]{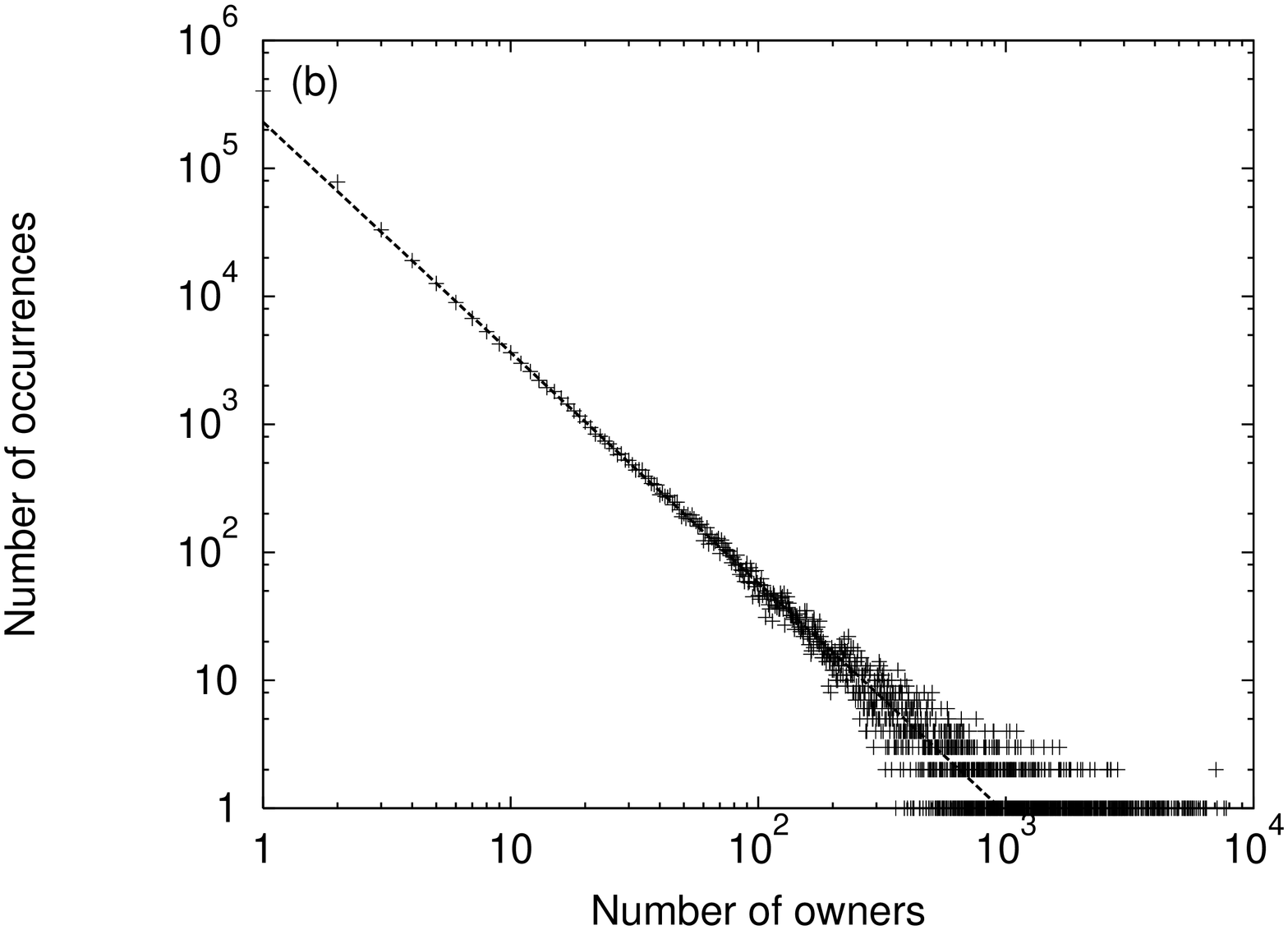}

\caption{\label{histo1}  (a) Histogram of  the number of music groups per user. 
The tail is fitted with the exponential $e^{-\frac{n_G}{150}}$ (dashed line), where $n$ is the number of links per node. A specific interval examined in the text is indicated
by vertical lines : average $<n_G>=140$, width 50; (b) Histogram for the audience 
size (number of listeners) per music group. This distribution behaves like the power-law $\sim n_L^{-1.8}$ }
\end{figure}

In the original data set, there are  $617900$ different music groups, although this value is skewed due to multiple (even erroneous) ways for a user to characterise an artist (e.g. {\em Mozart}, {\em Wolfgang Amadeus Mozart} and {\em Wolfgang Amedeus Mozart} count as three music groups) and $35916$ users.  There are 5028580  links in the bipartite graph, meaning that, on average, each user owns 140 music groups in his/her library, while each group is owned by 8 persons. For completeness, let us note that the listener with the most groups possesses 4072 groups ($0.6 \%$ of the total music library) while the group with the largest audience, {\em Radiohead}, has 10194 users ($28 \%$ of the user community). This asymmetry in the bipartite network is expected as users have in general specific tastes that prevent them from listening to any kind of music, while there exist {\em mainstream} groups that are listened to by a very large audience. 
This asymmetry is also observable in the degree distributions for the people and for the groups. The former distribution (see Fig.\ref{histo1}) is fitted respectively with an exponential $e^{-\frac{n}{150}}$ for large $n$ and the latter is a power-law $n^{-1.8}$, where $n$ is the number of links per node, i.e. $n_G$ or $n_L$ for groups and listeners respectively. Let us stress that such distributions are common in complex networks \cite{albert}. For instance, co-authorship networks also exhibit a bipartite asymmetry, and  power law  distribution $n^{-\alpha}$, with $\alpha \sim 2$ \cite{ramasco}.

Finally, let us mention the top ten groups in hierarchical order: {\em Radiohead,  Nirvana, ColdPlay, Metallica, The Beatles, Red Hot Chili Peppers, Pink Floyd, Green Day, Weezer} and {\em Linkin Park}. Obviously, the examined sample is oriented toward recent {\em rock} music. This fact has to be kept in mind, as it determines the mainstream music trend in the present sample, and could be a constraint on expected universality. This is left for further studies.

A common way to represent and to study bipartite networks consists in
projecting them onto links of one kind \cite{newman0}.
The standard projection method simplifies the system  to a unipartite network, where nodes are e.g. the listeners and where two listeners are connected if they have at least one music group in common. 
This scheme, that leads to a helpful representation in the case of collaboration networks \cite{newman00, newman1}, is unfortunately meaningless in the case under study. Indeed, due to the existence of mainstream music groups, the unipartite network is almost fully connected, i.e. most of the listeners are linked in the unipartite representation. For instance, {\em Radiohead} fully connects $28 \%$ of the user community whatever the rest of their music library contents.
This projection method definitely leads to an oversimplified and useless representation. We refine it by focusing on correlations between the users libraries. To do so, we define for each listener $\mu$ the $n_G$-vector $\overline{\sigma}^\mu$:
\begin{equation}
\label{vector}
\overline{\sigma}^\mu = (..., 1, ... , 0, ... ,1, ...)
\end{equation}
where $n_G=617900$ is the total number of groups in the system, $\mu \in [1, 35916]$ and where $\sigma^\mu_i=1$ if $\mu$ owns group $i$ and $\sigma^\mu_i=0$ otherwise.
This vector
is used as the {\em individual musical signature} (IMS), as mentioned in the introduction. 

In the following, we make a selection in the total number of users for computational reasons. To do so, we
have analysed a subset of $n_P=3806$ persons having a number of groups between
$[115, 165]$, -see Fig.\ref{filtering}, i.e.  around the average value $140$.  In order to quantify the correlations between two persons $\mu$ and $\lambda$, we introduce the symmetric correlation measure:
\begin{equation}
\label{cosine}
C^{\mu \lambda} = \frac{\overline{\sigma}^\mu . \overline{\sigma}^\lambda}{|\overline{\sigma}^\mu| |\overline{\sigma}^\lambda|} \equiv \cos \theta_{\mu \lambda} 
\end{equation}
where $\overline{\sigma}^\mu . \overline{\sigma}^\lambda$ denotes the scalar product between the two $n_G$-vector, and $||$ its  associated norm. This correlation measure, that corresponds to the cosine of the two vectors in the  $n_G$-dimensional space, vanishes when the persons have no common music groups, and is equal to $1$ when their music libraries are strictly identical.

\begin{figure}
\hspace{-0.7cm}
\includegraphics[angle=-90,width=3.6in]{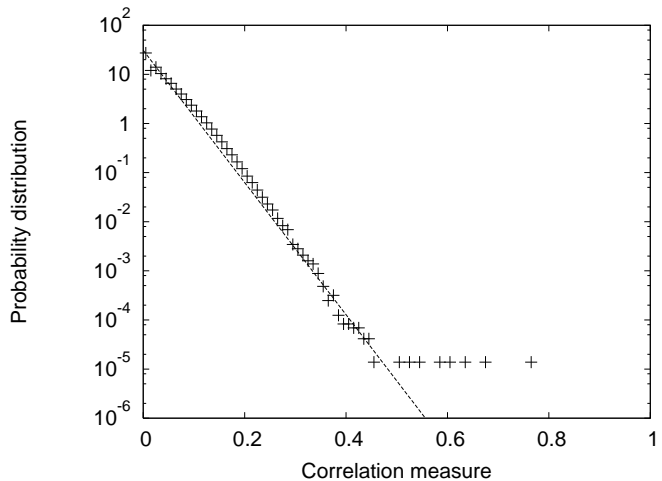}
\caption{\label{distC} Probability distribution of the matrix elements. The dashed line is the fitted exponential $\gamma e^{-\gamma C}$, with $\gamma = 31$. }
\end{figure}

At this level, the search for social communities requires therefore the analysis of the $n_P \times n_P$ correlation matrix $C^{\mu \lambda}$. A first relevant quantity is the distribution of the matrix elements $C^{\mu \lambda}$ that statistically characterises the correlations between listeners. Empirical results show a rapid exponential decrease of the correlation distribution (Fig.\ref{distC}), that we fit with $31 e^{-31 C}$, so that people in the sample are clearly
discriminated by their music taste, i.e. they are  characterised by non-parrallel vectors. This
justifies the use of his/her music library as a unique IMS of the listener.

\begin{figure}
\hspace{0.4cm}
\includegraphics[width=3.0in]{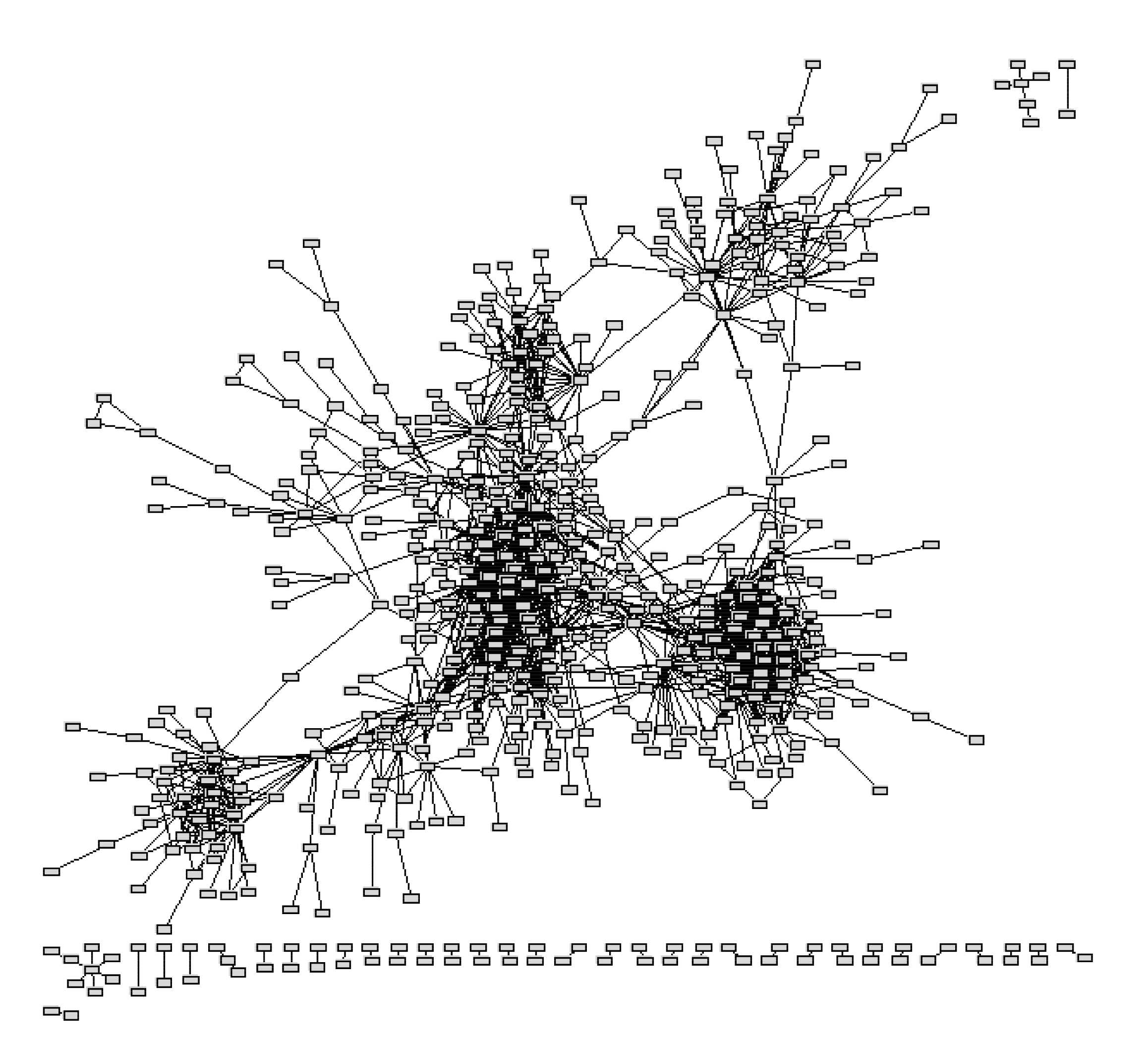}

\vspace{-0.1cm}
\includegraphics[width=3.0in]{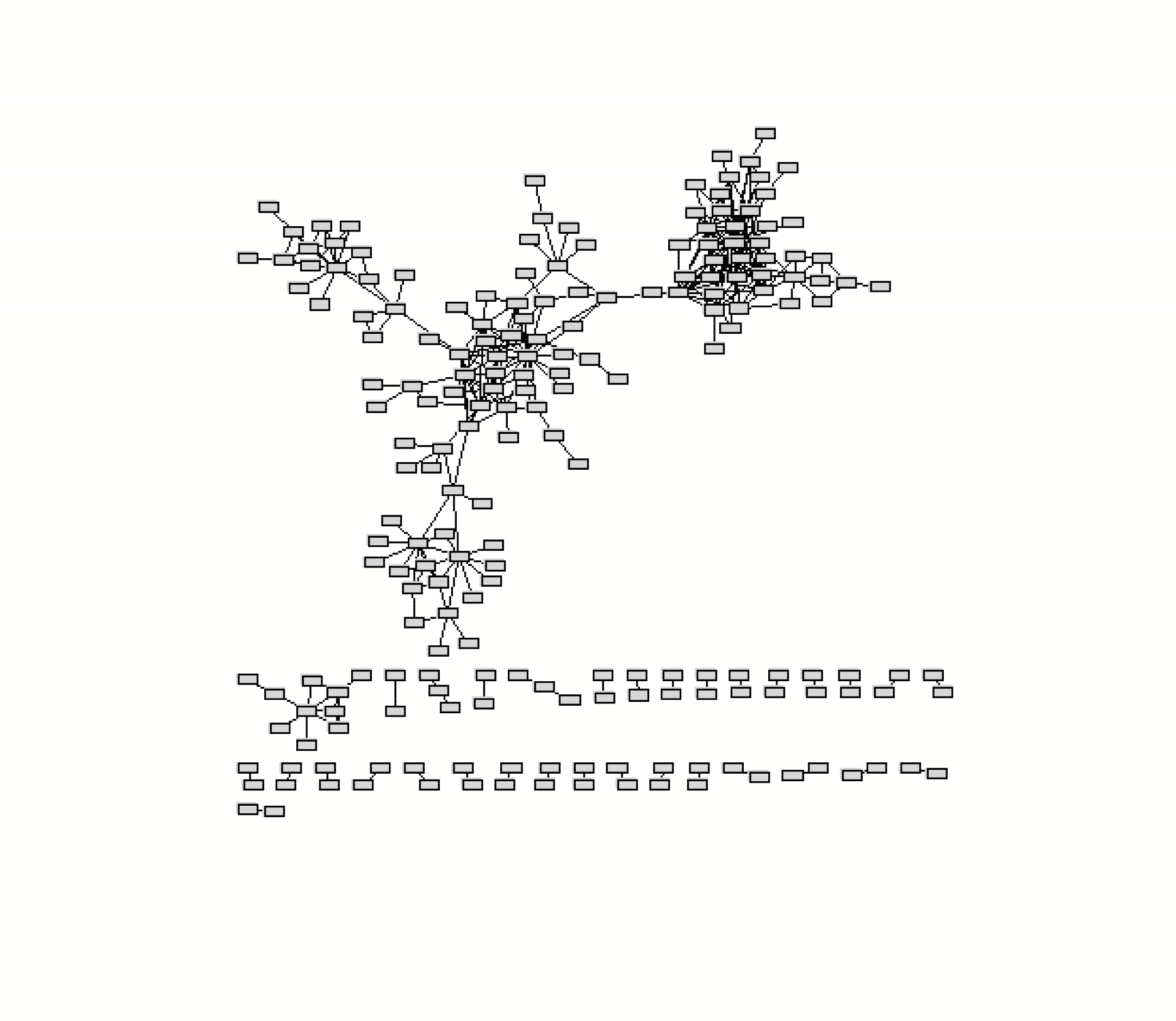}

\vspace{-1.5cm}
\includegraphics[width=3.0in]{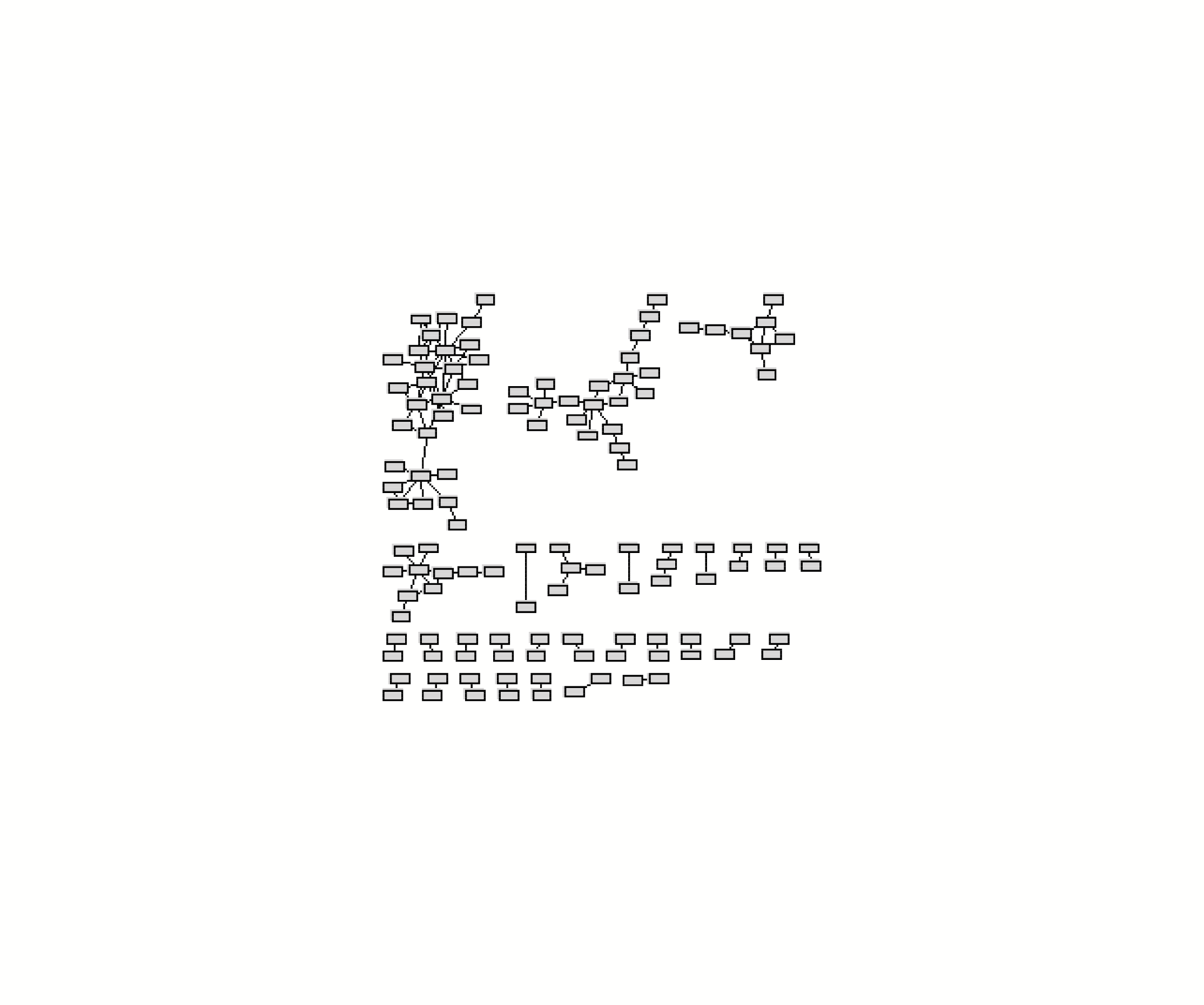}

\vspace{-1.8cm}
\caption{\label{filtering} Graph representation of the listener filtered correlation matrix for 3 values of the filter parameter $\phi=0.275$, $\phi=0.325$ and $\phi=0.35$, displayed from top to bottom. The graphs were plotted thanks to the {\em visone} graphical tools \cite{visone}.}
\end{figure}

\section{Percolation idea-based filtering}

\subsection{Listeners network}
\begin{figure}
\includegraphics[angle=-90,width=3.6in]{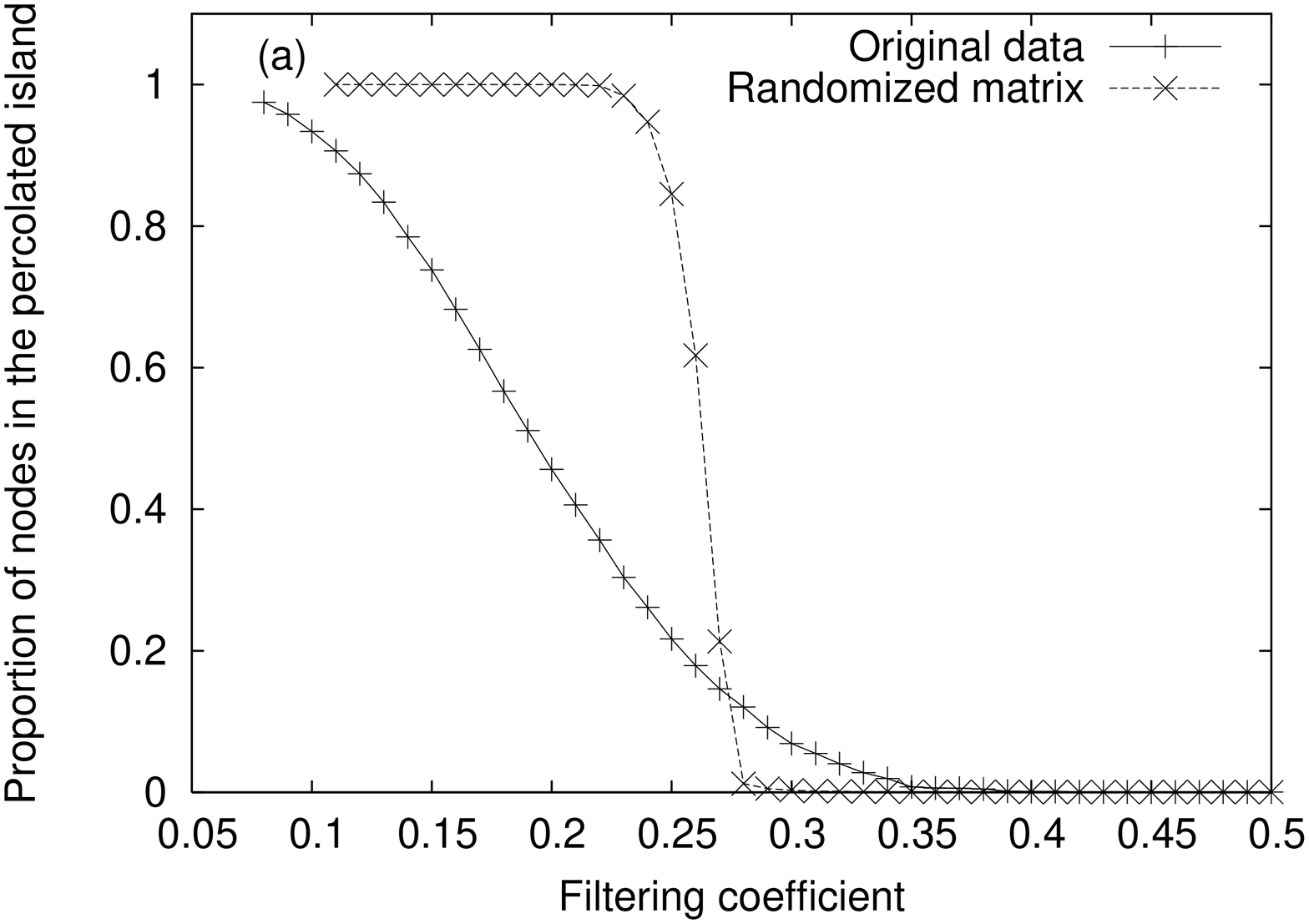}

\hspace{-0.5cm}
\includegraphics[angle=-90,width=3.6in]{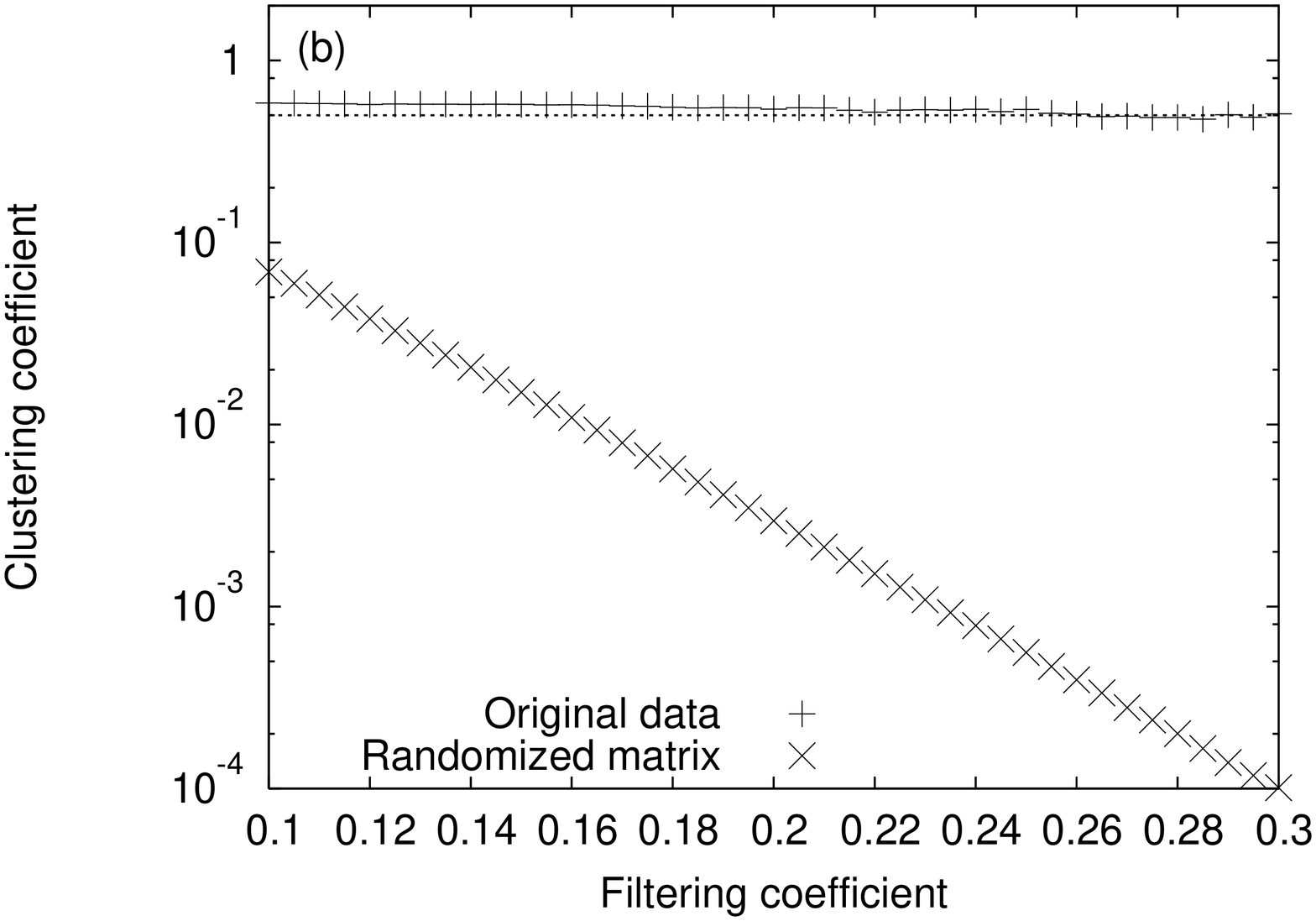}

\caption{\label{filter}  (a) Proportion of nodes in the percolated island as a function of the filtering coefficient $\phi$, for the original listeners correlation matrix $C^{\mu \lambda} $ and the corresponding randomised matrix;(b) Dependence of the clustering coefficient $C$ on the filtering coefficient $\phi$. The dashed line, at $C=0.5$, is a guide for the eye.}
\end{figure}

In order to extract communities from the  correlation matrix $C^{\mu \lambda}$, we use the following method. We define the filter coefficient $\phi \in [0,1[$, and filter the matrix elements so that $C_f^{\mu \lambda}=1$ if $C^{\mu \lambda}> \phi$, and let $C_f^{\mu \lambda}=0$ otherwise. In figure \ref{filtering}, we show the graph representation of the filtered matrix for increasing values of $\phi$. For the sake of clarity, we have only depicted the individuals that are related to at least one person, i.e. {\em lonely} persons are self-excluded from the network structure, whence from any community. One observes that, starting from a fully connected network, increasing values of the filtering coefficient remove less correlated links and lead to the formation of communities. These communities first occur through the development of  strongly connected components \cite{scc}, that are peninsulas, i.e. portions of the network that are {\em almost} disconnected from the main cluster,  themselves connected by inter-community individuals. Further increasing the filtering coefficient value leads to a removal of these inter-community individuals, and to the shaping of well-defined islands, completely disconnected from the main island. 
Let us stress that this systematic removal of links is directly related to percolation theory.  It is therefore interesting to focus on the influence of the network structuring along percolation transition ideas.
To do so, we compare the bifurcation diagram of the empirical data with that obtained for a randomised matrix, i.e. a matrix constructed by a random re-disposition of the elements $C^{\mu \lambda}$. As shown in Fig.\ref{filter}a, the correlated structure of the network broadens  the interval of the transition as compared to the uncorrelated case. Moreover, the correlations also seem to displace the location of the bifurcation, by requiring more links in order to observe the percolation transition. This feature may originate from  community structuring that restrains network exploration as compared to random structures \cite{lambi}.

As a first approximation, we restrict the scope to the formation of islands in the matrix analysis, i.e. to the simplest organised structures.  From now on, we therefore associate the breaking of an island into sub-islands to the emergence of a new sub-community, and,  pursuing the analogy, we call the largest connected structure the {\em mainstream community}.
Before going further, let us stress that the projection method described above is exactly equivalent to that of \cite{newman0} when $\phi=0$.

\begin{figure}
\includegraphics[width=3.2in]{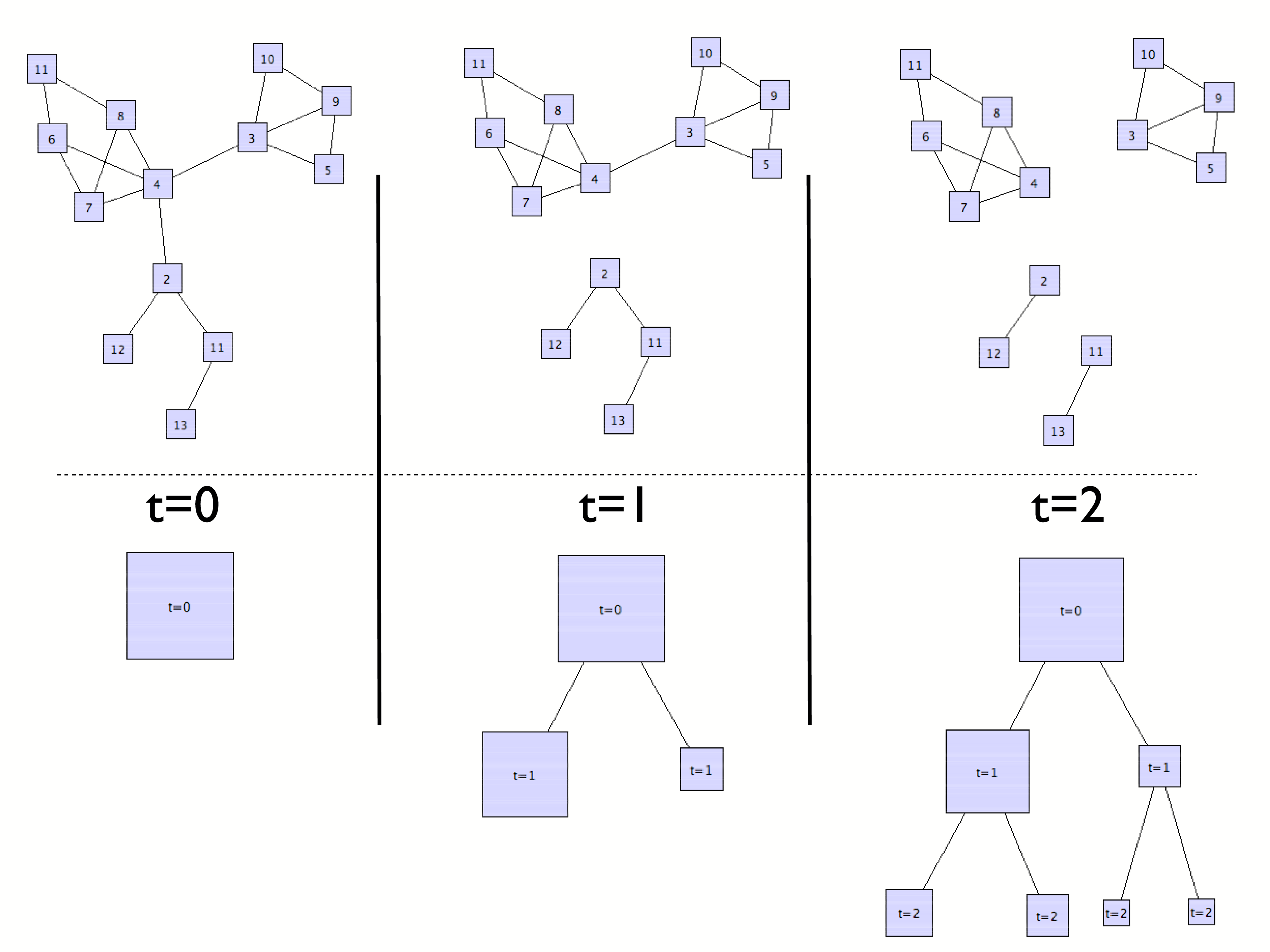}
\caption{\label{explication}  Branching representation of a correlation matrix. At each increasing step (t=0,1,2) of the filter $\phi$, links are removed, so that the network decomposes into isolated islands. Theses islands are represented by  squares, whose size depends on the number of nodes in the island. Starting from the largest island, branches indicate a parent relation between the islands. The increasing filter method is applied until all links are removed.}
\end{figure}

\begin{figure}
\includegraphics[width=3.2in]{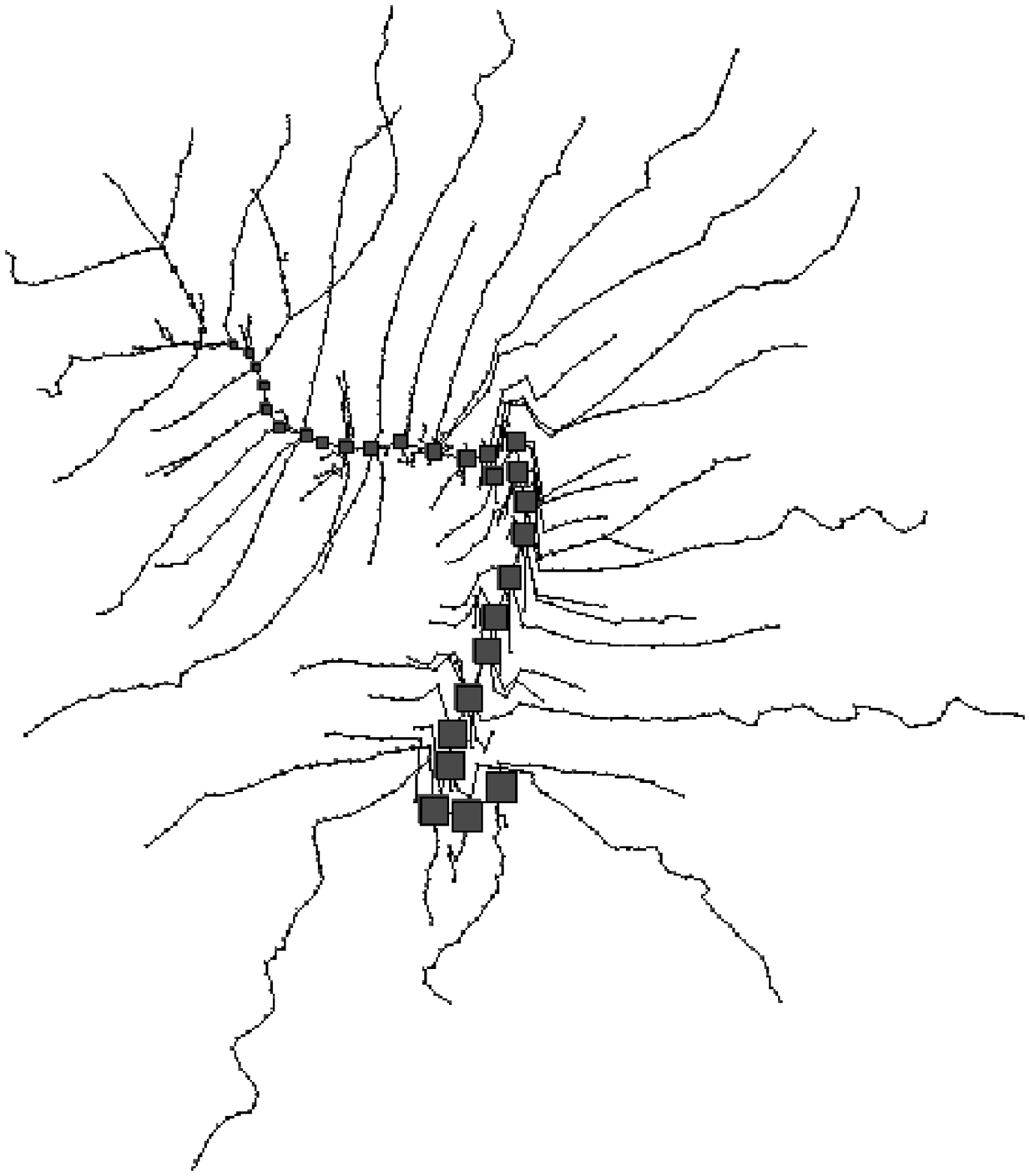}

\includegraphics[width=3.2in]{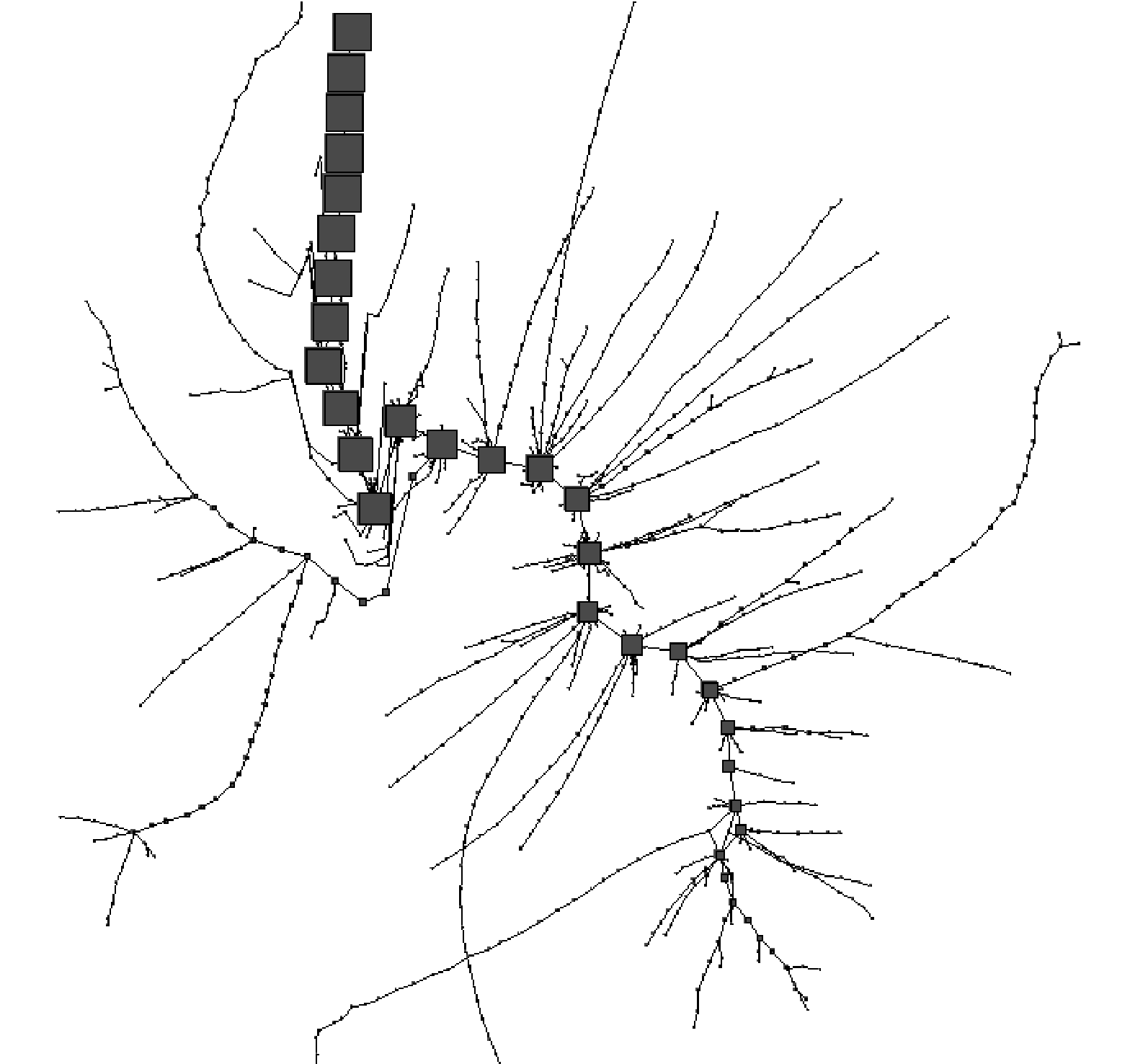}

\caption{\label{tree}  In the upper figure, branching representation of the listener $C^{\mu \lambda}$ correlation matrix. The filtering, with parameter ranging from $0.2$ to $0.5$ (from bottom to top) induces a snake of squares at each filtering level. The shape of the snake as well as its direction are irrelevant. In the lower figure, branching representation of the music groups correlation matrix, the filtering parameter ranging from $0.3$ to $0.6$ (from top to bottom).}
\end{figure}

In the following, we use a branching representation of the community structuring (see Fig.\ref{explication} for the sketch of three first steps of an arbitrary example). To do so, we start the procedure with the lowest value of $\phi$, here $\phi=0.2$, and we represent each isolated island by a square whose surface is proportional to its number of  listeners. Then, we increase slightly the value of $\phi$, e.g. by 0.01, and we repeat the procedure.  From one step to the next step, we draw a bond between emerging sub-islands and their parent island. The filter is increased until all bonds between nodes are eroded (that is, there is only one node left in each island). Applied to the above correlation matrix $C^{\mu \lambda}$ (Fig.\ref{tree}), this representation leads to a compact description of the series of graphs as those found in Fig.\ref{filtering}. Moreover, the  snake structure gives some insight into the diversification process by following branches from their source toward their extremity.
 The longer  a given branch is followed, the more likely it is forming a well-defined community.
  
 In order to focus on collective effects, we have studied in detail the behaviour of the clustering coefficient \cite{newman0}, that is a measure of the density of triangles in a network, a triangle being formed every time two of one node's neighbours are related between them.
This quantity is a common way to measure social effects in complex  networks, and measures, roughly speaking,  whether the friend of a friend is a friend. In figure \ref{filter}b, we plot the dependence of this quantity $C$ vs. the filtering coefficient $\phi$. Moreover, in order to highlight the effects of correlations, we compare the results with those obtained for the above randomised matrix. 
Our analysis shows a very high value of $C$, almost $\phi$ independent for the original matrix. This suggests that the way people acquire their music taste is a highly social mechanism, likely related to its identification role as described in the introduction. 

\subsection{A Typical individual Music Signature}
Before focusing on the genre-fication of music groups,  we give here,  as an empirical example, the music library of one person. This list is intended to indicate the diversity of groups that characterise a listener, as well as his/her community. We write in bold the music groups that are common to his/her sub-community, found by the PIB technique, and in normal characters those that are owned only by the individual. There are $117$ different music groups.

{\bf     Music library}: {\footnotesize \em {Air+ {\bf  New Order}+ Jane's Addiction+ DJ Krush+ {\bf  Massive Attack}+ {\bf  DJ Shadow}+ {\bf   Beastie Boys}+ Orbital+  {\bf   Blur}+  {\bf    Pixies}+ Leftfield+ {\bf    Sonic Youth}+ {\bf  David Bowie}+ Primus+ {\bf    Jeff Buckley}+ The Smiths+ {\bf  Daft Punk}+ {\bf   Joy Division}+ {\bf   Smashing Pumpkins}+ {\bf   Chemical Brothers}+ Korn+ {\bf  Eminem}+ {\bf  Nirvana}+ Radiohead+ {\bf   Grandaddy}+ Travis+ {\bf   Oasis}+ PJ Harvey+ {\bf   Manic Street Preachers}+ Roots Manuva+ Unkle+ {\bf  Linkin Park}+ Atari Teenage Riot+ {\bf  Kula Shaker}+ The Police+ James Iha+ {\bf     Semisonic}+ {\bf   Weezer}+ Anastacia+ {\bf  Rob Dougan}+ {\bf   Eels}+ Fatboy Slim+  {\bf    Green Day}+      {\bf    Lostprophets}+      {\bf    System of a Down}+ U.N.K.L.E.+ El-P+      {\bf    Bee Gees}+      {\bf    Duran Duran}+      {\bf    Therapy?}+ The Prodigy+      {\bf    Foo Fighters}+      {\bf    JJ72}+ Alkaline Trio+      {\bf    The Beatles}+      {\bf    Incubus}+ Prodigy+      {\bf    Muse}+ And You Will Know Us By The Trai+      {\bf    Jimmy Eat World}+      {\bf    Ash}+      {\bf    Rival Schools}+ Cher+ At The Drive-In+ Johnny Cash+ Mansun+ Queens of the Stone Age+      {\bf    Basement Jaxx}+ Dave Matthews Band+ Dj Tiesto+ Cast+      {\bf    The Strokes}+ Anthrax+ Ian Brown+      {\bf    Saves The Day}+ Morrissey+ Police+ Modest Mouse+ Interpol+ St Germain+      {\bf    The Beach Boys}+ Bonnie Tyler+ Theme+      {\bf    Fenix*TX}+      {\bf    Snow Patrol}+ The Cooper Temple Clause+ Buddy Holly+      {\bf    Nada Surf}+ onelinedrawing+ Michael Kamen+ Remy Zero+ Ernie Cline+      {\bf    Quicksand}+ Olivia Newton John+      {\bf    Polar}+ Ikara Colt+ Keiichi Suzuki+ Rivers Cuomo+ Paddy Casey+      {\bf    Billy Talent}+ Mireille Mathieu+ Jack Dee+ Tomoyasu Hotei+ Daniel O'Donnell+ Hope Of The States+      {\bf    Franz Ferdinand}+ The Shadows+ THE STILLS+      {\bf    The RZA}+ The Mamas and the Papas+ Melissa Auf Der Maur+ Barron Knights+      {\bf    The Killers}+      {\bf    R.E.M.}+     {\bf    Jay-Z   DJ Danger Mouse}+ Pras Michel Feat ODB and Maya+ The Monks Of Roscrea}}

Obviously, this person belongs to a music community characterised by a mixture of the usual music genres, including Pop/Rock, 80's Pop, Electro, Alternative... This eclecticism  indicates the inadequacy of such music subdivisions to characterise individual and  collective listening habits.

\subsection{Music groups network}
In view of the above, it is interesting to  introduce a new way to build  music sub-divisions, i.e. based upon the listening habits of their audience. 
 To do so, we have applied the PIB approach to a sample composed of the top 5,000 most-owned groups. This limited choice was motivated by the possibility to identify these groups at first sight. Each music group is characterised by its  signature, that is a vector:
 \begin{equation}
\overline{\gamma}^i = (..., 1, ... , 0, ... ,1, ...)
\end{equation}
of $n_L$ components, where $n_L=35916$ is the total number of users in the system, and where $\gamma^i_{\mu}=1$ if the listener $\mu$ owns group $i$ and $\gamma^i_{\mu}=0$ otherwise. By doing so, we consider that the audience of a music group, i.e. the list of persons listening to it,
 identifies its signature,  as we assume that the music library characterises that of  an individual.
 
\begin{figure}
\includegraphics[width=3.2in]{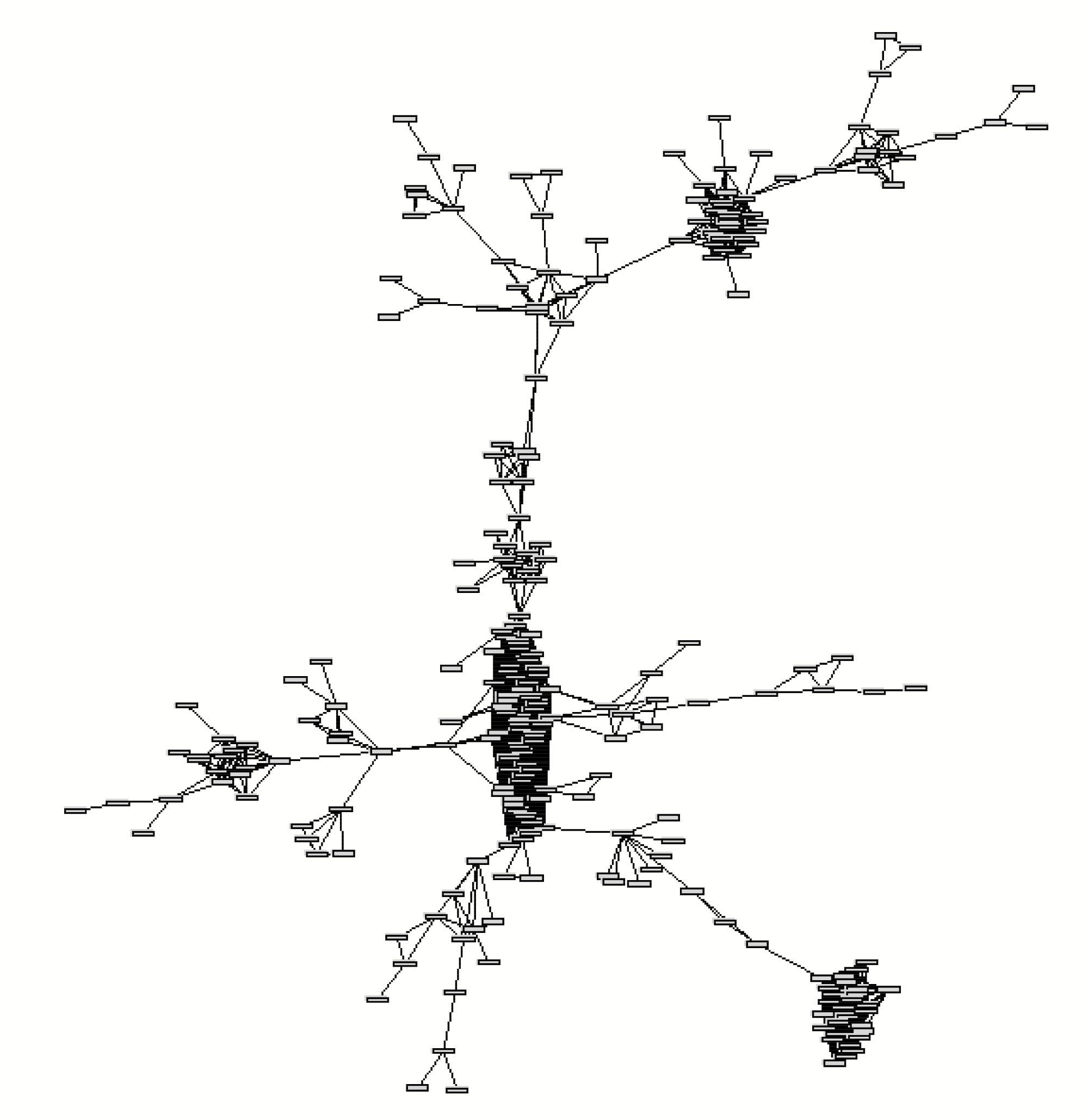}

\includegraphics[width=3.2in]{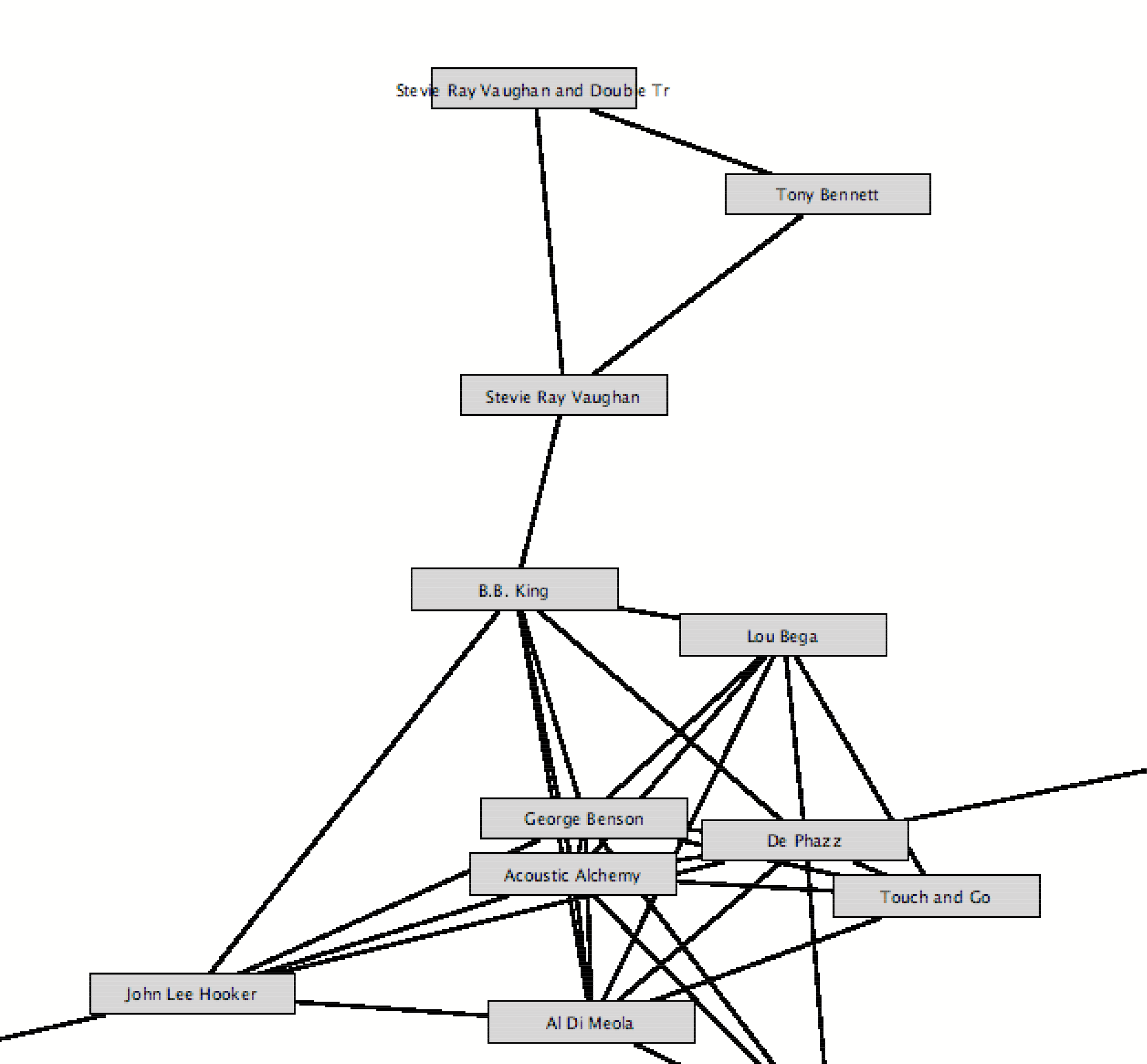}

\caption{\label{percoMusic} In the upper figure, typical percolated island of music groups for $\phi=0.45$. It is composed of 247 nodes and 4406 links. In the lower figure, zoom on a small structure of the percolated island, that is obviously composed of {\em guitar heroes}, e.g. {\em B.B. King}, {\em S.R. Vaughan}, {\em A.D. Meola}... Let us also note that {\em S.R. Vaughan} appears through two different ways that are linked by our analysis.} 
\end{figure}

   The next step consists in projecting the bipartite network onto a unipartite network of music groups. To do so, we 
build the correlation matrix for the music groups as before, and filter it with increasing values of the filtering coefficient. As previously, the action of filtering erodes the nodes, thereby revealing a structured percolated island (Fig. \ref{percoMusic}) that breaks into small islands.
The resulting tree representation of the correlation matrix (Fig. \ref{tree}) shows clearly long persisting branches, thereby suggesting a high-degree of common listenership. 
Poring over the branches of the top $5000$ tree \cite{gideon2}, we find many standard, homogenous style groupings. Amongst these homogeneous cliques, there are
{\footnotesize \em {[George Strait, Faith Hill, Garth Brooks, Clint Black, Kenny Chesney, Shania Twain, Alan Jackson, Martina McBride, Alabama, Tim McGraw, Reba McEntire, Diamond Rio, John Michael Montgomery, SheDaisy, Brooks and Dunn, Clay Walker, Rascal Flatts, Lonestar, Brad Paisley, Keith Urban],~~~
[Kylie Minogue, Dannii Minogue, Sophie Ellis Bextor],~~~
[Serge Gainsbourg, Noir D\'esir], ~~~
[Billie Holiday, Glenn Miller, Benny Goodman],~~~
[Morrissey, Faith No More, Machine Head, The Smiths, Rammstein, Smashing Pumpkins, Slipknot, Tomahawk, Mr. Bungle]}}, that are country, dance pop, geographically localised i.e. France, swing jazz and rock groupings respectively.

In contrast, many of the islands are harder to explain from a standard genre-fication point of view. In some cases, the island may be homogeneous in one music style, but show some unexpected elements, like:
{\footnotesize  \em {[Spain In My Heart (Various), The Pogues, Dave Brubeck Quartet, Crosby, Stills, Nash and Young, Phil Ochs, Billy Bragg, Clem Snide, Sarah Harmer, Mason Jennings, Kirsty MacColl, tullycraft, Ibrahim Ferrer, Sarah Slean, Penguin Cafe Orchestra, Pretenders, Joe Strummer and The Mescaleros, Freezepop] }}
that is a folk/folk  cluster, with odd members like the Brubeck Jazz Band, for example. But other groupings defy monolithic style categorisation, like:
{\footnotesize  \em {[The Jon Spencer Blues Explosion, Yello, Galaxie 500, Prince and the Revolution, Ultra Bra, Uriah Heep, Laurent Garnier],
[Crosby, Stills, Nash and Young, Orb, Zero 7, Royksopp, Stan Getz]}}.
The latter include unexpected mixtures of  Indie Rock/Funk/Hard Rock/Dance, and Folk/Electro/Jazz respectively. 

Consequently, the PIB approach  reveals evidence of unexpected collective listening habits, thereby uncovering trends in music.
As a matter of fact, these anomalous entities have been shared by multiple listeners. This seems to confirm the role of collective listening habits in the individual building of music taste.
It is important to note that the PIB method neglects the relevance of  the main island structuring by identifying "music genres"/"listener communities" with isolated islands. It is obviously a drastic simplification that may lead to the neglect of pertinent structures, and therefore requests a more detailed exploration of the network structure.

\begin{figure}
\hspace{-0.2cm}
\includegraphics[angle=-90,width=3.50in]{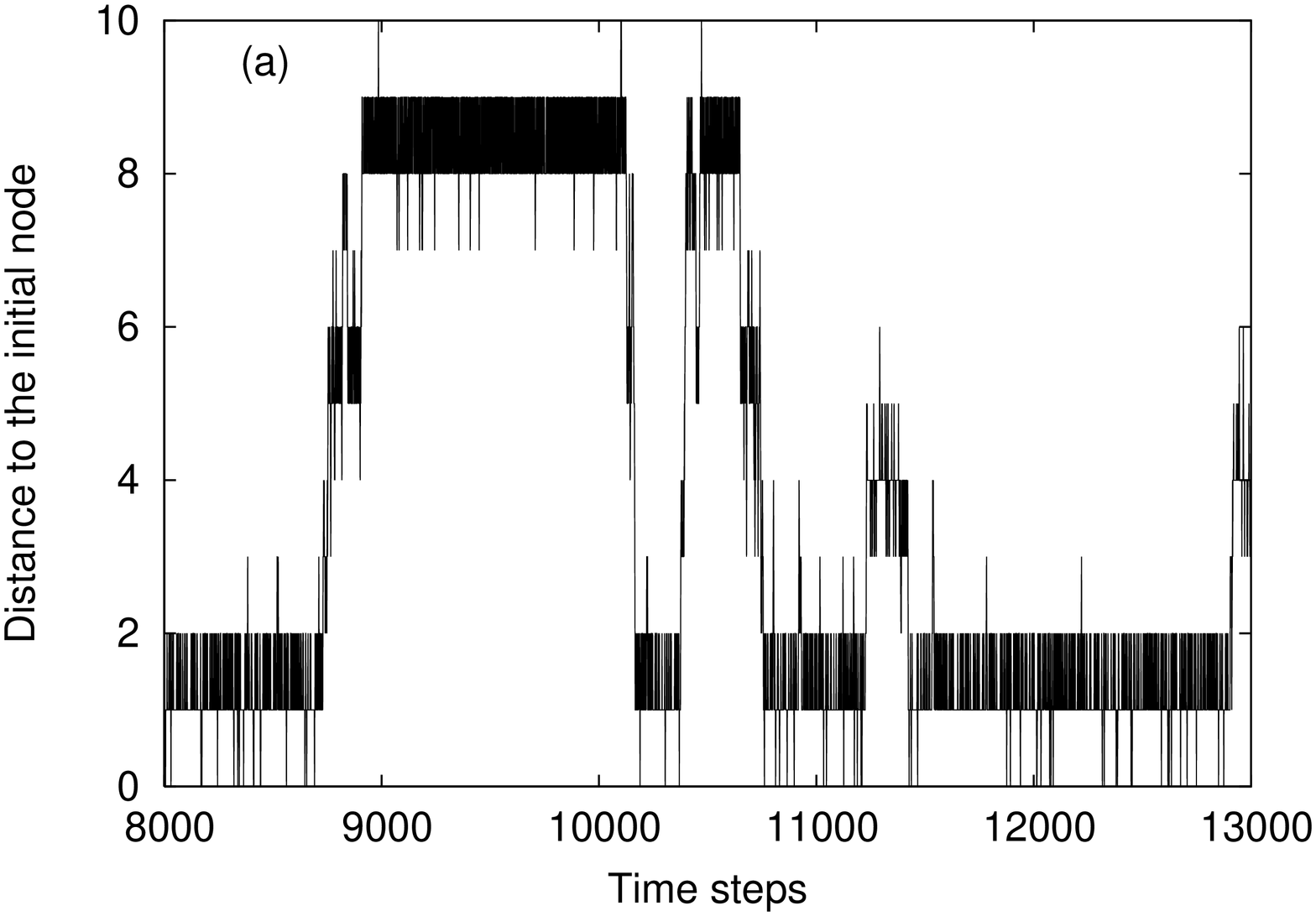}

\hspace{-0.8cm}
\includegraphics[angle=-90,width=3.50in]{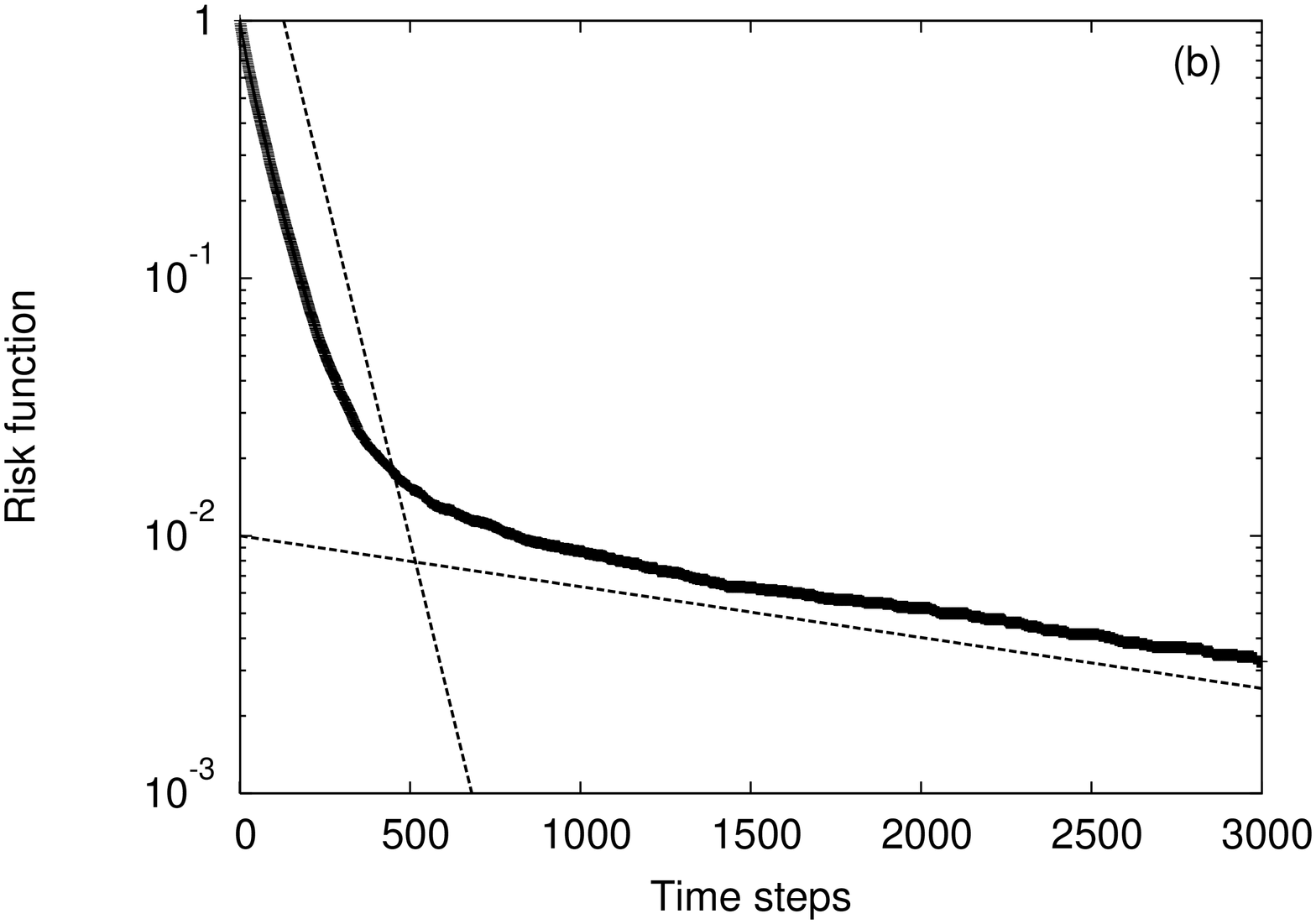}

\caption{\label{rw}  (a) 
Time evolution of the distance to the initial node during the RW on the network of Fig.7. The network exploration exhibits clearly the passage from one cluster to another, followed by long stopovers in the latter cluster. (b) Risk function $R(t)$ of the signal. The dashed lines are guides for the eye, and represent the exponential relaxations $e^{-t/80}$ and $e^{-t/2200}$. }
\end{figure}

\section{Random walk exploration}

In this section, we consider an alternative method for revealing the internal structures of the network. The method is based on a random walking  exploration of the percolated island. The random walk (RW) starts at some node, i.e. the {\em initial node}. At each time step, we choose randomly one of its links, and move the walker to the connected node. Moreover, we keep track of the distance from the occupied node to the original node $d_{0i}$ as a function of time $i$.  By definition, the distance between two nodes is the length of the smallest path between them. 
 The initial node is chosen to be 
the central node of the percolated island, namely the node $c$ that minimises the average distance with other nodes in the island:
\begin{equation}
<d>=\frac{1}{(n_I-1)} \sum_{i\neq c}^{n_I} d_{ci}
\end{equation}
 where $n_I$ is the number of nodes in the island. 
 
 In the following, we focus on the percolated island of figure \ref{percoMusic}, that is composed of
$n_I=247$ nodes, and $4406$ links. The percolated island clearly exhibits penininsulae, that link alike music groups. For instance, the cluster in the centre of the figure is "hard rock" oriented, with music groups like {\em The killing tree}, {\em Unearth}, {\em Murder by Death}... This is also illustrated in the lower graph of figure 7, where we zoom on a small structure that encompasses {\em guitar heroes}, e.g. {\em B.B. King}, {\em S.R. Vaughan}, {\em A.D. Meola}, {\em G. Benson}...
In the case under study, the central node is the music group {\em Murder by Death}, that is located in the hard rock cluster.

The resulting time series (Fig.\ref{rw}a), that is directly related to the subjacent path geometry, seems to indicate the existence of different time-scales, associated with the large-scale structures in the network. In order to analyse the time series,
we have focused on the probability of return toward the initial node. To do so, we have measured the time intervals $\tau$ between two passages of the walker on the initial node, and calculated the distribution $f(\tau)$ of these time intervals. Moreover, in order to study the rare events associated to the tail of the distribution, we focus on the risk function $R(t)=\int_t^\infty f(\tau) d\tau$. By construction $R(0)=1$ and $R(\infty)=0$. The results, that are plotted in figure \ref{rw}b, clearly reveal two time scales: a rapid time scale (80 time steps) that determines the internal dynamics in one cluster; a slow time scale (2200 time steps) that characterises the passage from one cluster to another one.
Let us stress that detrended fluctuation analysis of the random walk \cite{lambi3} leads to the same conclusion.
 
 \bigskip

\begin{figure}
\includegraphics[angle=-90,width=3.50in]{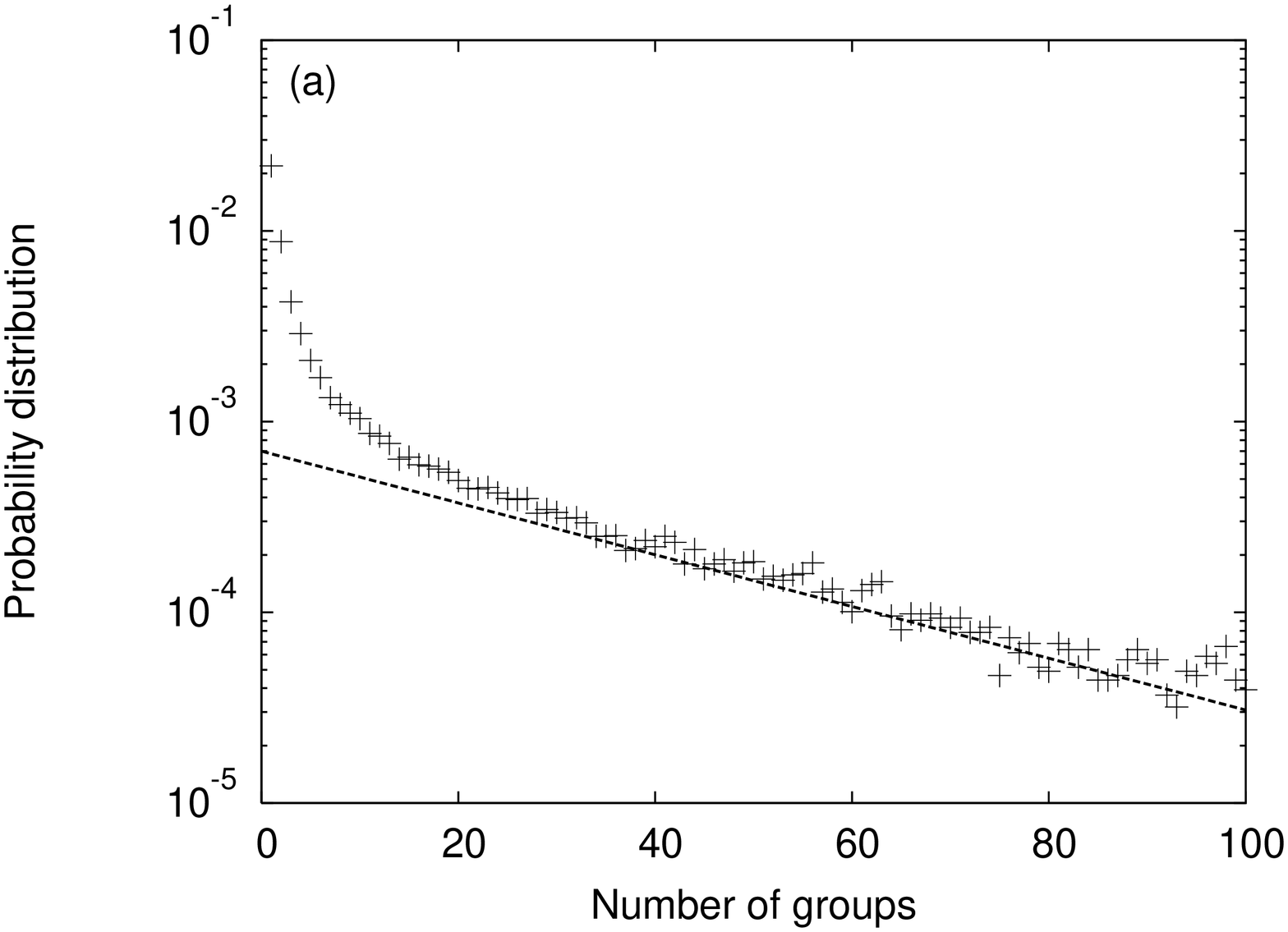}

\includegraphics[angle=-90,width=3.50in]{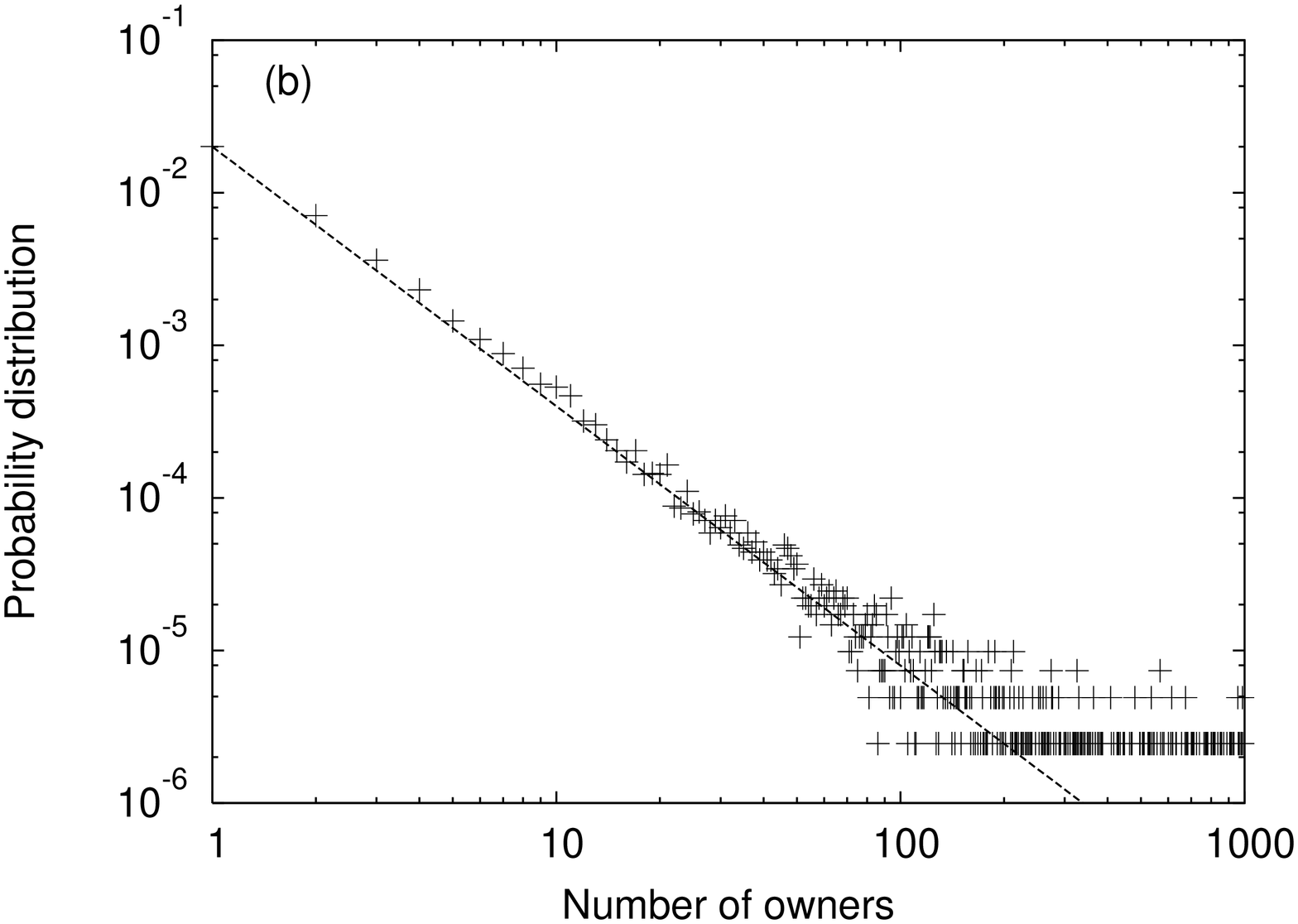}

\caption{\label{distSim}  Simulation results of PICIM for parameters given and explained in the text : (a) 
distribution of  the number of music groups per user
; (b) distribution of the audience per group. The dashed line is the power-law $\sim n^{-1.7}$.}
\end{figure}

\section{personal identification - community imitation model}
The empirical results of the previous section suggest that
a person's musical taste derives from an interplay between Personal Identification, i.e. his/her individual choice, and Community Imitation, i.e. the collective trend. In order to test this assumption, we introduce the PICI model where personal music libraries build through two processes. On one hand,  collective effects, originating from social interactions between individuals, are mimicked by an exchange of music groups between {\em similar} individuals. In order to define this similitude between two persons, we compare their music libraries, and favour the pair interactions between people having alike music taste, as in a Potts model \cite{potts}.  On the other hand, there are individual mechanisms that push people to distinguish themselves from their community. We model such a dynamics by individual random choices.  We neglect the effect of an external field, like advertising, on an individual behaviour.
Moreover, in order to reproduce the observed degree distributions of the bipartite graph \cite{ramasco}, we assume that the networks are growing in time. This is done in a way that music groups are chosen with preferential attachment \cite{albert}, i.e. with a probability simply proportional to their audience.

These requirements are put into form as follows.
The system is composed by $L(t)$ users and $M(t)$ music groups, that are initially randomly linked. At each (Monte Carlo) time step, three processes may occur:

(i) A new user may enter the system, with probability $p_I$. His/her library contains one music group, chosen randomly in the set of previous groups with preferential attachment. 

(ii) A randomly picked user adds a new music group to the library, with probability $p_N$. This new group appends to the collection of available music in the system.

(iii) Two randomly chosen users exchange their music knowledge, with probability $p_E$. The pair is selected with a probability proportional to $e^{\frac{(\cos \theta_{\mu \lambda} -1)}{T}}$, where $\theta_{\mu \lambda}$ is the angle between the vectors of their music libraries (Eq.\ref{vector}), defined by their cosine (Eq.\ref{cosine});  the {\em temperature} $T$ is a parameter that represents the ability of qualitatively different communities to mix together. If the pair is selected, we compare the two music libraries, and give to each user a fraction of the unknown groups of his/her partner. Let us stress that this rule ensures preferential attachment for the music groups.

Some representative results of the simulations obtained from the model are selectively shown in Fig.\ref{distSim}, for a typical simulation set, with $p_I=0.02$, $p_N=0.03$, $p_E=0.03$ and $T=0.13$.  A complete analysis of the PICI model phase space variables and the dynamics will be  presented elsewhere. The simulations were stopped after 200 time steps/node, in a system composed by 22800 users, 15126 music groups and 442666 links. 

The degree distributions of the bipartite graph are depicted in Fig.\ref{distSim}. The results reproduce quite well the exponential and the power-law features experimentally found (Fig.\ref{histo1}). For the group distribution, the exponent is close to the empirical value $1.8$. Moreover, different simulations show that this value remains in the vicinity of $2$ for a large set of parameter values.  

For the user distribution, simulations also reproduce the deviations from the exponential for small number of groups $n_G$, as observed in Fig.\ref{histo1}. We have noticed (unshown) that these deviations diminish for increasing values of $T$. 
This uncovers that the self-organising mechanisms associated to community structuring are responsible for the extreme deviations. 

The dependence of the clustering coefficient $C$ on the filtering coefficient  has also been considered.  It is found that the the simulations reproduce qualitatively well the almost constant high value of $C$ found in Fig.\ref{filter}. However this behaviour ceases to be observed for large values of the temperature, i.e. in systems where collective effects do not develop by construction of the model. This seems to confirm the crucial competing roles played by individual choices and community influence in order to reproduce the observed data.

\section{Conclusion}

In this article, we study empirically the musical behaviours of a large sample of persons. Our analysis is based on complex network techniques, and leads to the uncovering of individual and collective trends from the data. To do so, we use two methods. On one hand, we use percolation idea-based techniques that consist in filtering correlation matrices, i.e. correlations between the listeners/music groups.  Moreover, the communities/music genres are visualised by a branching representation.
On the other hand, we explore the structure of the main percolated island by randomly walking the network.  
The goal is to map its internal structure and correlations onto a time series, that we analyse with standard statistical tools. 

The method allows to reveal non-trivial connections between the listeners/music groups. It is shown that if some empirical sub-divisions respect the standard genre classification, many sub-divisions are harder to explain from a standard genre-fication point of view.
These collective listening habits, that do not fit the neat usual genres defined by the music industry, represent the {\em non-conventional} taste of listeners. They could therefore be an alternative {\em objective} way to classify music groups. 
These collective genre-hopping habits also suggest a growing eclecticism of music listeners \cite{gideon2}, that is driven by curiosity and self-identification, in opposition to the uniform trends promoted by commercial radios and $Major$ record labels \cite{margolis}. 

We would like to point that the above methods should help finding and visualising structures in a large variety of networks, e.g. the detection/classification of trends in marketing, show business and financial markets. 
Applications should also be considered in taxonomy \cite{taxonomy}, in scientometrics, i.e. how to classify scientific papers depending on their authors, journal, year, keywords..., and in linguistics \cite{lingui}.
From a more theoretical point of view, this work is closely related to the theory of hidden variables \cite{agatka, boguna,masuda}, i.e. the hidden variables being here some intrinsic property of the music groups \cite{lambb}, and
should provide an empirical test for this theory.

Whence, we  introduce a simple grow model, that reproduces quite well the results obtained from the empirical data, i.e. the observed degree distributions of the networks.
It is important to point out  that the ingredients of the model are very general, i.e. imply competition between personal identification and community imitation (PICI). Consequently, PICI should apply to a larger variety of systems than the music networks hereby investigated, but also to other networks
 such as collaboration networks in science \cite{ramasco}. In a statistical physics sense, the model contains  Potts model-like ingredients for opinion and taste formation.

\begin{acknowledgments}
Figures  3, 5, 6 and 7 were plotted thanks to the {\em visone} graphical tools \cite{visone}.
R.L. would like to thank especially T. Padilla for providing data of musicmobs, and A. Scharnhorst for fruitful suggestions. We are indebted to  G. D'Arcangelo who explored the music groupings associated to Fig.3, and  shared his findings  with us prior to publication \cite{gideon2}. This work 
has been supported by European Commission Project 
CREEN FP6-2003-NEST-Path-012864.
\end{acknowledgments}

\end{document}